\documentclass[10pt,letterpaper]{article}

\makeatletter
 \let\Ginclude@graphics\@org@Ginclude@graphics 
\makeatother

\usepackage{amsmath,amssymb}
\usepackage{changepage}
\usepackage{textcomp,marvosym}
\usepackage[numbers,sort]{natbib}
\usepackage{nameref,hyperref}
\usepackage[right]{lineno}
\usepackage[nopatch=eqnum]{microtype}
\usepackage{natbib}
\usepackage{setspace} 
\usepackage[font=scriptsize]{caption}
\usepackage[aboveskip=1pt,labelfont=bf,labelsep=period,justification=raggedright,singlelinecheck=off]{caption}
\usepackage{lastpage,fancyhdr,graphicx}
\usepackage{epstopdf}

\makeatletter
\renewcommand{\@biblabel}[1]{\quad#1.}
\makeatother

\begin{document}

\begin{flushleft}
{\Large
\textbf\newline{Connect-while-in-range: modelling the impact of spatial constraints on dynamic communication network structures}
}
\newline
\\
Niek Kerssies*\textsuperscript{1,2},
Jose Segovia-Martin\textsuperscript{1,3},
James Winters\textsuperscript{4}
\\
\bigskip
\textbf{1} School of Collective Intelligence, University Mohammed VI Polytechnic, Rabat, Morocco
\\
\textbf{2} Department of Sociology, Rijksuniversiteit Groningen, Groningen, The Netherlands
\\
\textbf{3} Complex Systems Institute of Paris Ile-de-France, Centre national de la recherche scientifique, Paris, France
\\
\textbf{4} Centre for Culture and Evolution, Department of Psychology, College of Health, Medicine and Life Sciences, Brunel University London, London
\\
\bigskip

*niek.kerssies@um6p.ma

\end{flushleft}
\section{Abstract}
Like other social animals and biological systems, human groups constantly exchange information. Network models provide a way of quantifying this process by representing the pathways of information propagation between individuals. Existing approaches to studying these networks largely hypothesize network formation to be a result of cognitive biases and choices about who to connect to. Observational data suggests, however, that physical proximity plays a major role in shaping the formation of communication networks in human groups. Here we report results from a series of agent-based simulations in which agents move around at random in a bounded 2D space and connect while within communication range. Comparing the results to a non-spatial model, we show how including spatial constraints impacts our predictions of network structure: ranged networks are more clustered, with slightly higher degree, higher average shortest path length, a lower number of connected components and a higher small-world index. We find two important drivers of network structure in range-constrained dynamic networks: communication range relative to environment size, and population density. These results show that neglecting spatial constraints in models of network formation makes a difference for predicted network structures. Our simulation model quantifies this part of the process of network formation, realized by simply situating individuals in an environment. The model also provides a tool to include spatial constraints in other models of human communication, as well as dynamic models of network formation more generally. 

\section{Introduction}
Observational data suggests that physical proximity plays a major role in shaping the formation of social and communication networks in human groups \citep{cowgillUsingPredictionMarkets2009, stopczynskiHowPhysicalProximity2018, tothInequalityRisingWhere2021a}. Current experiments and models in human groups, however, focus mostly on static network structures chosen as conditions by researchers \citep{momennejadCollectiveMindsSocial2021, centolaNetworkScienceCollective2022, centolaSpreadBehaviorOnline2010, centolaSpontaneousEmergenceConventions2015, moserInnovationfacilitatingNetworksCreate, derexPartialConnectivityIncreases2016, derexDivideConquerIntermediate2018, lazerNetworkStructureExploration2007, bavelasCommunicationPatternsTask1950}. When dynamic networks are considered, network evolution is hypothesized to be shaped by individual connection biases, strategies or preferences \citep{conteStrategiesSocialNetwork, almaatouqAdaptiveSocialNetworks2020, nakayamaSocialInformationSpontaneous2019}. Relatedly, models, experiments and fieldwork in cultural evolution have postulated "model-based" biases and strategies with which humans and other animals select others to observe, communicate and interact with \citep{kendalSocialLearningStrategies2018, boydCultureEvolutionaryProcess1985, lalandSocialLearningStrategies2004, canteloupWildPrimatesCopy2020, kendalChimpanzeesCopyDominant2015}. 

For cognitive hypotheses on individual choices and biases like these to explain network formation, a simple condition needs to hold: each individual needs to be able to choose from the whole population under consideration at every point in time. In many natural settings, this condition does not obtain. To name just a few examples of network 'pre-selection', humans and other animals inherit social contacts from their parents \citep{smithNaturePrivilegeIntergenerational2022, mulderIntergenerationalWealthTransmission2009, cantorInterplaySocialNetworks2013,  ilanyRankdependentSocialInheritance2021}; those with access to university education have access to many more social contacts \citep{bokanyiAnatomyPopulationscaleSocial2023}; and parental income and status impact properties of offspring social networks such as diversity of contacts \citep{pikettyCapitalTwentyFirstCentury2015, gemarParentalStatusConnection2024}. Social and communication networks are not shaped by individual choice alone, but prefigured partially by social as well as physical circumstances.

Physical proximity is a circumstance that is both very simple and very fundamental. Communication is an activity that takes place in an environment and using a physical communication medium with a functional range. Nonhuman animals employ various media such as chemical, visual, tactile, and seismic signals, while humans additionally use technologies ranging from conversations to the Internet \citep{langbauerElephantCommunication2000, kotaInsectCommunication2022, acerbi2020cultural}. Data from digital communications technology has greatly expanded our understanding of human communication networks \citep{backstrom2011anatomy, onnelaStructureTieStrengths2007, bokanyiUrbanHierarchySpatial2022a}. However, many interactions remain face-to-face, as they have throughout much of human (evolutionary) history and in other social species. In these settings, physical proximity and location in the environment continue to play a much larger role. Moreover, the structure of online social networks is still shaped significantly by geography and existing real-world social ties \citep{acerbi2020cultural, backstrom2011anatomy, bokanyiUrbanHierarchySpatial2022a}. Whether the group in question is a school of fish, a robot swarm, or a human crowd, part of the collective computation of communication network structure is not perfomed by individual decisions, but by the physical constraints of the environment and ranged communication. 

In this paper, results are reported from a series of agent-based simulations designed to capture and quantify this dynamic. Agents randomly move through a 2D coordinate space and connect to others only while within a specified communication range. The resulting network structures are compared to results from a model without positions and movement in which agents connect at random. Like a recent model by Chimento and Farine \citep{chimentoContributionMovementSocial2024}, the model simulates the impact of movement through a 2D space on network structure and information transmission. Also like that model, we consider statistics of dynamic networks at each timestep, rather than properties of cumulative interaction graphs (static networks composed of interactions combined over time). Unlike that model, in which population size and communication range stay constant, we consider the complete meaningful parameter space of communication range and population density, mapping the influence of these parameters on 6 network structure measures, from $N$ disconnected components to fully connected networks. Our model is also like a cellular automata model by Vining and colleagues \citep{viningHowDoesMobility2019} where agents move randomly through a 2D space and connect within range, though they focus on the impact of spatial constraints on performance on a consensus task, rather than network structure outputs.

We find that introducing location, random movement and ranged communication introduces much more clustered networks, with slightly higher degree, higher average shortest path length, a lower number of connected components and a higher small-world index than a dynamic non-spatial model. All of these differences are more extreme when range is low. Our model shows how simply locating communicating individuals in an environment changes our predictions of the communication network structures that can form, and provides a tool for including spatial constraints in models of (communication) network formation. In so doing, the model also functions as an agent-based alternative to existing network formation algorithms, creating networks from bottom-up interactions between nodes rather than setting top-down global node properties.

\section{Materials and methods}
\subsection{Model}
\subsubsection{Random movement and the range rule.}

In agent-based models, each of a population of agents follows an update rule every timestep. In this model, the model parameters are \emph{N}, \emph{g} and \emph{r}. \emph{N} is the population size, \emph{g} is the grid size such that \emph{g * g} is the area of the coordinate space, and \emph{r} is agent communication range. \emph{N} agents are initialized and assigned a unique random coordinate value \emph{(x,y)} within the \emph{g * g} grid.

Each timestep, every agent (1) moves, (2) connects, and (3) disconnects. To move, an agent chooses from a uniformly random distribution of its adjacent integer coordinates, including its current location. In other words, it selects a tile at random in its Moore neighborhood. To keep agents within the bounded space and to prevent agents sharing the same position, tiles within the neighborhood with coordinate values below 0 or exceeding \emph{g}, and tiles currently occupied by other agents, are excluded from the random selection of a new position. If all tiles in the neighborhood are out of bounds or occupied, the agent remains in the current position.

\begin{figure}[!h]
\includegraphics[width=\linewidth]{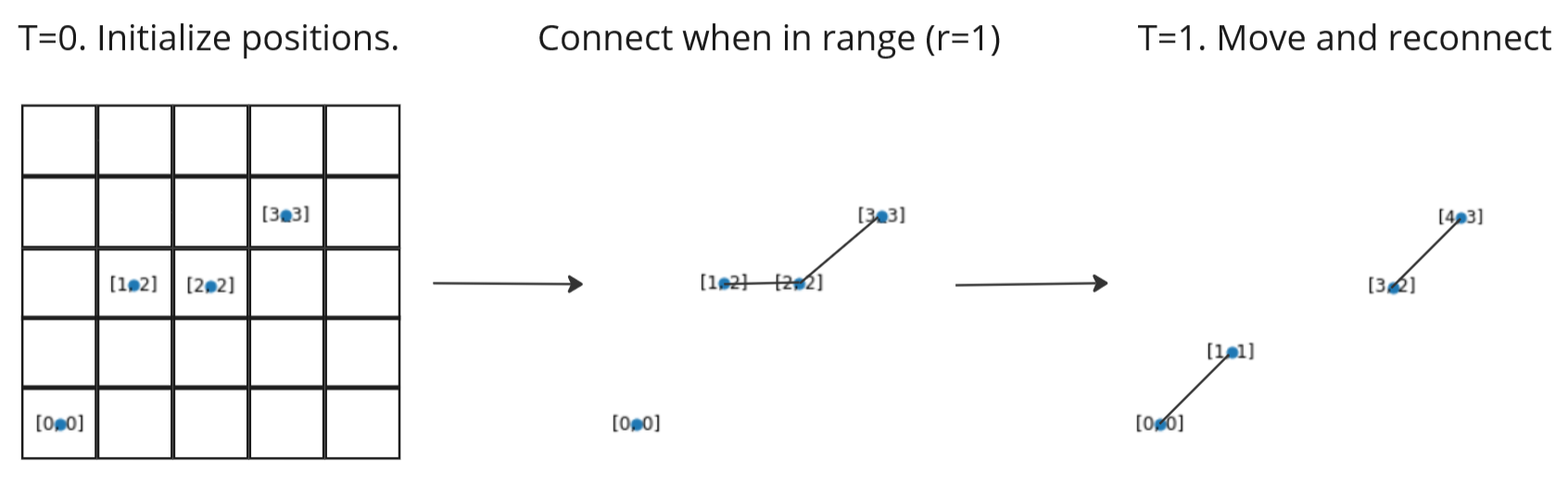}
\caption{{\bf Demonstration of a possible initialization and timestep in the model for $ \bf{N=4, r=1, g=4.}$} Agents are randomly assigned a unique (x,y) coordinate. The coordinate space is graphically represented here as a grid(left), and agents are shown with their coordinate values. At every timestep, every agents follows the range rule: if the Euclidean distance between the agents is less than or equal to \emph{r}, a link is created between  them. Thus, from the initial position shown on the grid, the graph shown in the middle of the figure is formed. At the start of the next timestep, agents move at random and follow the range rule again. When previously in range agents are now out of range, the link is cut. Hence, from every agent performing random movement and following the range rule, a new network is formed. }
\label{fig1}
\end{figure}

To connect, after moving, each agent creates an undirected network link with all other agents that are currently in range. Range is calculated as Euclidean distance in the coordinate space, rounded to integer values for x and y. Euclidean distance uses the Pythagorean theorem that for right-angled triangles, $c^2 = a^2+b^2$ (where \emph{a} and \emph{b} are the base and altitude, and \emph{c} the hypothenuse of the triangle). Agent \emph{i} checks whether agent \emph{j} is in range by putting the difference in coordinates $x_j-x_i$ and $y_j - y_i$ as \emph{a} and \emph{b}, and calculating the distance to \emph{j} as \emph{c}; so $r = \sqrt{ (x_j-x_i)^2 + (y_j - y_i)^2}$. If the result is less than or equal to the value of \emph{r}, agent \emph{j} is within range. Finally, to disconnect, the agent deletes any links with agents that are no longer in range. The order in which agents are updated is randomized each timestep. Fig. \ref{fig1} shows the initialization and first timestep of a possible model run schematically.

\subsubsection{Null model.}
The intention of the simulations reported in this paper is to demonstrate the work being done by spatial constraints on communication network formation. In order to do so, the results are compared to a non-spatial model. In the non-spatial model, each timestep agents simply connect to each other agent with probability \emph{P(connect)} which is set globally as the same for all agents. In order to compare this model's outputs to that of the range model, \emph{N} can be set to the same value or varied in the same range of values, while \emph{P(connect)} corresponds to \emph{r}. Like \emph{r}, the \emph{P(connect)} parameter can be varied through its whole meaningful range of values, from isolated agents to fully connected networks. A connection probability of 0 will mean \emph{N} unconnected nodes, while a probability of 1 will always create fully connected networks. The parameter \emph{g} has no equivalent in the non-spatial model and there are no coordinates assigned to agents. Finally, in order to make the non-spatial model a dynamic process as well, a random disconnect chance is included, which is set to $1-P(connect)$. This way, the probabilities of connecting and disconnecting mirror those of the range model, as the area not covered by an agent's range is equal to the inverse of the area covered (for a calculation of connection probability under random movement and ranged communication in an unbounded space, see \citep{viningHowDoesMobility2019}).

\subsection{Test setup}
Simulations consisted of several tests. In a test, one of the model's three input parameters \emph{N,g,r} was chosen to vary while the others remained constant. For each integer value of the varying input parameter, both the range and the null model ran for 100 rounds of 100 timesteps, where at each timestep, all agents were updated in a random order and output measures were collected.  

\begin{figure}[!h]
\includegraphics[width=\linewidth]{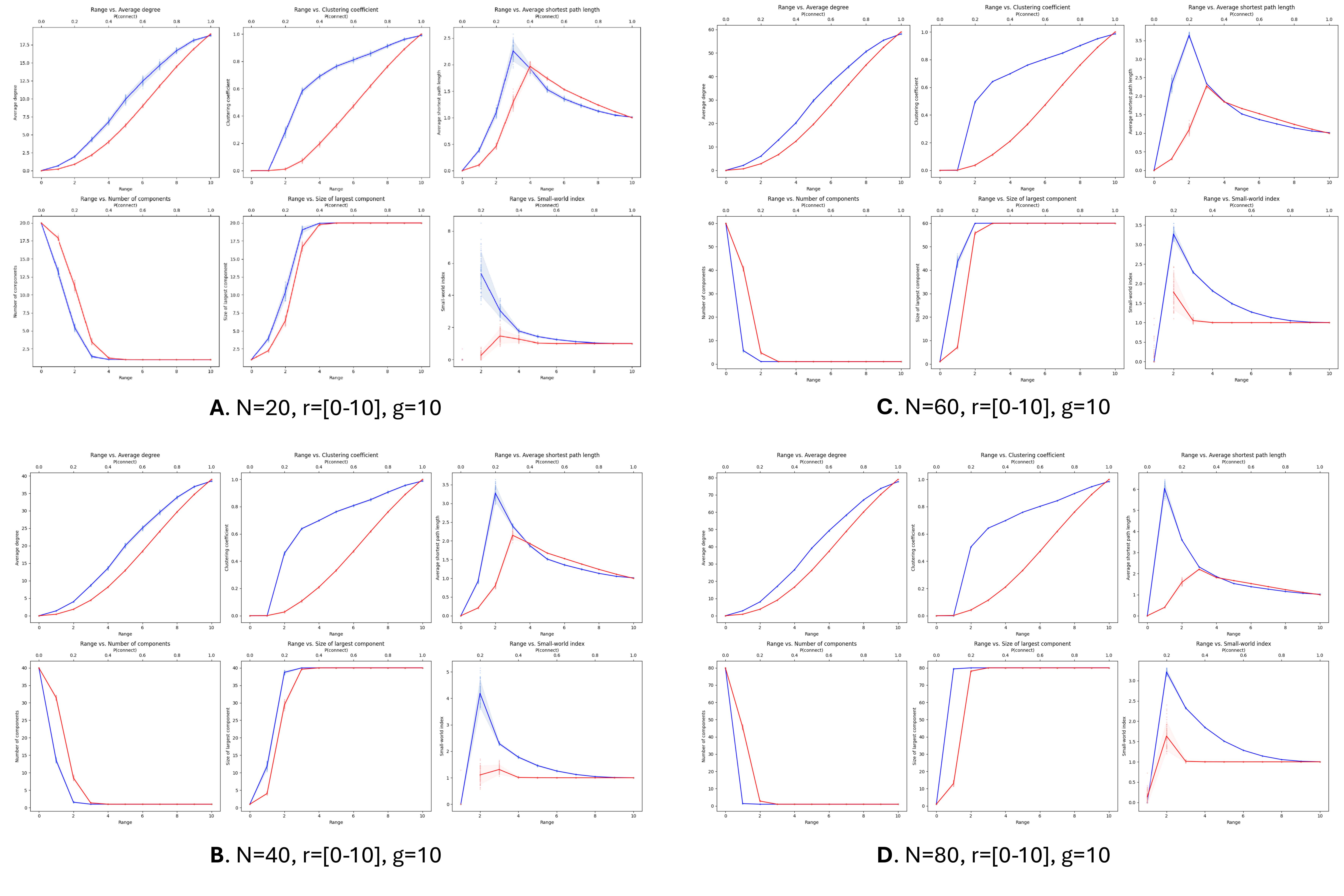}
\caption{{\bf Network properties of range model (blue) and null model (red) networks for varying range between 0 and 10 (x-axes) within runs and varying population density 0.2,0.4,0.6, and 0.8 between runs (top to bottom)}. Dots in the graphs are time-averages of 100 timesteps, of which 100 rounds are collected at each input setting of \emph{r}. Lines are drawn through the means of these time averages, and transparent areas show 1.5 standard deviations. Most visibly at low range settings (when spatial constraints are most articulated), networks are much more clustered, have slightly higher degree, higher average shortest path length, a lower number of connected components and a higher small-world index than the null model. As \emph{N} varies between tests, most network measures stay the same, being predominantly influenced by \emph{r} in the range model and \emph{P(connect)} in the null model. At lower range settings, however, average shortest path length (and as a result, small-world index) increases noticeably with increasing \emph{N}, as the disconnected groups that form at lower range settings become larger while still being highly clustered.}
\label{fig2}
\end{figure}

For the first set of tests, range was varied between \emph{r=0} and $r=10$ while \emph{N} was kept at constant values, varying between tests from $N=1$ to $N=100$ (see Fig. \ref{fig2} for results for $N=20, 40, 60$, and 80). Varying range up to $r=g$ ensures that in terms of network structure, the complete meaningful range of \emph{r}-values is tested. At $r=0$, no nodes are connected; at $r=g$, where the diameter of each agent's range covers the whole length of the coordinate grid, networks are always fully connected. In the null model, the connection probability parameter \emph{P(connect)} was varied between 0 and 1 with intervals of 0.1, similarly ensuring that the simulations start at disconnected nodes and end at fully connected networks.

In the second set of tests, in each test \emph{N} in both the range and the null model was varied between 1 and 49, while $g=7$ and \emph{r} varied at constant values between tests, from \emph{0} to \emph{7}. \emph{N} varies up to 49 because when $g=7$ there are $7*7=49$ unique positions in the environment. Since agents avoid sharing the same position, as \emph{N} increases towards $g x g$, the space becomes increasingly "crowded" until it is completely saturated and agents remain in place, each in a unique position. For this parameter, then, the meaningful range is determined by population density $N/(g * g)$, varying between $1/49$ at $N=1$ and $49/49=1$ at $N=49$. Since it lacks spatial positions, the null model has no analog to \emph{g}. Therefore we show results here only for varying \emph{N}; for varying \emph{g}, see the supplemental materials.  In the null model, \emph{N} was varied over the same values while \emph{P(connect)} was set to $r/g$. This way the null model mimics the connection probability of the range model, in which an agent's probability to connect is proportional to the range relative to the environment size.

\subsection{Output measures}
Each timestep, 6 network properties were calculated in both models. The network properties calculated are average degree, clustering coefficient, average shortest path length, number of connected components, size of largest component, and small-world index.

Networks (sometimes called graphs) are mathematical objects represented by dots (nodes or vertices ) and lines (links or edges) between the dots. Degree is a property of single nodes: the amount of connections it has in the network. The average degree \emph{K} is simply the population average of node degrees $k_i$: $ K = \frac{\sum_{i=1}^{N} k_i}{N}$. In this model, nodes do not connect to themselves and redundant connections are not possible. Therefore, the maximum amount of connections one node can have is $N-1$. For a population of 10 nodes, average degree will therefore be a value between 0 and 9. Note that if the population average degree is equal to $N-1$, this indicates a complete graph: the network where each node is connected to all other nodes. 

Clustering indicates the fraction of a node's network neighbors that are themselves connected to each other. To illustrate, say node A is connected to three others B, C and D. B and C are connected, but C and D are not and neither are B and D. In this case, ABC forms a fully connected triangle, while ABD and ACD do not. Clustering can be indicated as the ratio of actual to possible fully connected triangles, in this case $1/3$. This is the clustering coefficient. This coefficient is calculated for the whole network each timestep, resulting each time in a value between 0 and 1. Note that a value of 1 for this coefficient corresponds to the complete graph at which average degree is $N-1$; if all possible triangles are fully connected, then the whole network is fully connected. 

The shortest path length between two nodes is amount of links that need to be traversed in order to get from one agent to the other. Average path length is the average taken over all node pairs. Here, too, an average value of 1 means that the networks are fully connected, since it indicates that all pairs have a direct path of 1 link between them. 

In this model, the agents will not always form a single network. More often, especially at low range (and at low connection probability in the null model), the population will consist of several mutually unconnected local networks, or single isolated nodes. We calculated the number of such connected components and the size of the largest connected component. When the number of components is 1, the population forms a single connected graph; when it is equal to \emph{N}, no nodes are connected. Conversely, when the largest component is equal to $N$, the population forms a single connected graph, and when it is 1, no nodes are connected. Note that a population that forms a single connected network, where there is at least one path between all nodes, is not necessarily a fully connected network, where there is a direct link between all nodes.

Finally, we calculated the small-world-index of networks. Strictly speaking, a small world network is one defined by the parameters \emph{n,k} and \emph{p}, where \emph{n} is the population size, \emph{k} is the number of neighbors on a ring lattice that each node connects to (therefore each node has the same degree \emph{k}), and \emph{p} is the probability that a link is "rewired": replaced by a link to a node chosen at random from the whole population \citep{wattsCollectiveDynamicsSmallworld1998}.  A direct comparison between a small-world networks and range model networks, however, does not make much sense, for two reasons. First, the range model is a spatial model: in a small-world network, nodes have a fixed position on a ring lattice, while in the range model, nodes move around through a 2D coordinate space. Second, in a small-world network \emph{k} specifies a global degree value. When rewiring probability \emph{p} is 0, this means that all nodes have degree \emph{k}. This is very rare in a dynamic, spatially explicit model with random movement; it can only occur when agents happen to form an equidistant ring in space.

In order to still have an indication of the occurrence frequency of small-world-like structures, there is the small-world index \citep{humphriesNetworkSmallWorldNessQuantitative2008}. The small-world index \emph{S} of a graph \emph{G} is the ratio of \emph{C} to \emph{L}. \emph{C} is equal to $CG/CR$, where \emph{CG} is the clustering coefficient of \emph{G}, and \emph{CR} is the clustering coefficient of a random graph with the same number of \emph{n} nodes and \emph{m} edges as \emph{G}. Similarly, \emph{L} is equal to $LG/LR$, where \emph{L} is the average shortest path length. \emph{S}, then, is calculated as $S = (CG/CR)/(LG/LR)$. This measure captures the feature of small-world graphs that most nodes are locally connected, while some are rewired to form 'longer' ties outside of local clusters. Adding only a few of such long ties connects relatively clustered local communities, creating much shorter average shortest path lengths while keeping clustering high \citep{wattsCollectiveDynamicsSmallworld1998}. If, therefore, clustering is low and average shortest path length is high compared to a random model, \emph{S} will be low; if the opposite is the case, \emph{S} is high.

\section{Results}
\subsection{Ranged communication creates much more clustered graphs than a non-spatial model.}

The most conspicious difference between the range model and the null model is the difference in clustering. At lower values of \emph{r}, the average clustering coefficient of networks increases sharply with increasing \emph{r}. The non-spatial model behaves in almost opposite fashion, networks with higher \emph{P(connect)} showing an initially slow and eventually steeper increase in clustering. 

Part of this difference in clustering is accounted for by the difference in average degree. As communication range and connection probability increase, agents form more connections, leading to higher population average degree. As more of the possible connections between agents are made, more connected triangles form, increasing the clustering coefficient. In the non-spatial null model, this effect seems to account for the increase in clustering, which closely follows the increase in degree. 

\begin{figure}
\includegraphics[width=\linewidth]{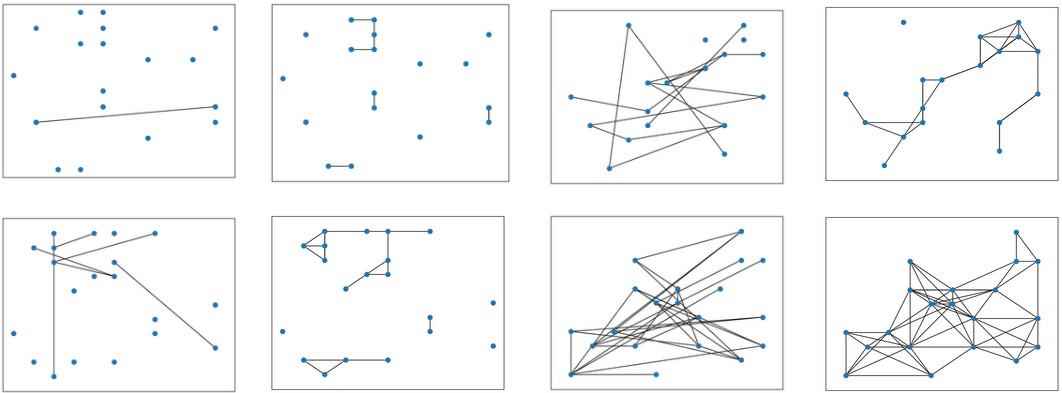}
\caption{{\bf Four sample network timesteps comparing the null model (left) to the range model (right), $N=18, g=10$}. Upper left: $r=1, pc=0.1$. Lower left: $r=2,pc=0.2$. Upper right: $r=3,pc=0.3$. Lower right: $r=4, pc=0.4$.The spatial layout created by the positions of the agents in the range model graphs are copied in the null model comparisons. At similar connection probabilities, agents in the range model only connect to closer agents, creating more connected triangles.}
\label{fig3}
\end{figure}

In the range model, however, this straightforward relationship between degree and clustering is not observed, especially at lower values of \emph{r}. This is because there is an additional effect increasing clustering, introduced by range-constrained network formation. In the non-spatial model, the formation of closed triangles is purely a matter of chance: namely, the chance that an agent A at one timestep \emph{t} is connected to both agent B and C (or has been at a timestep before \emph{t}, and has not since disconnected), and that additionally B and C are connected at \emph{t}. Hence, degree and clustering simply increase as this connection probablity \emph{P(connect)} increases. In the range model, there is a condition of physical proximity to network links. Since movement is random, the probability of two agents A and B being within range and therefore connected is also well-modelled as a random process, where connection probability is proportional to the percentage of the area covered by an agent's range \citep{viningHowDoesMobility2019}. For any two connected agents A and B, B needs to be within the area covered by the communication range of agent A. For a third agent C, however, to connect to A, it needs to be within this area as well. For randomly moving agents, this means the space of possible coordinates drawn from in the random movement step is reduced from the entire space $g*g$ to the agent range area $2\pi r$, which is significantly smaller when $r:g$ is low. This means that if both B and C are connected to A, they have a much higher chance (than agents not connected to A) to connect to each other as well. Therefore, networks become more clustered simply because of introducing spatial communication constraints. As range increases, this requirement becomes less strict, as the likelihood that agent C in this example is close to agent B decreases as \emph{r} tends to \emph{g} and the area covered by a single agent's range increases proportionally. Fig. \ref{fig3} illustrates this point with network visualizations.

This same effect accounts for a lower average shortest path length in the range model. In the null model, any node can connect to any other node, generally creating 'well-mixed' networks with no preference for specific nodes. In the range model, there is a preference for connecting to agents close in the coordinate space, and as a result, a tendency to form local clusters. 'Long ties' that bridge these locally clustered communities are more rare and only happen on the edges of spatially clustered communities. As range increases, the clustering effect decreases while at the same time the formation of between-community ties becomes easier. This creates a distinct 2 phases to average shortest path length with increasing range. In the first phase, from $r:g=0$ to $r:g=0.2$, networks have increasingly higher average shortest path length, as distinct clustered groups form. After $r:g=0.2$, average shortest path lenghts start to decrease, as the clustering effect diminishes and longer connections form between the groups, until fully connected networks are reached at $r=g$ and average shortest path length becomes 1. These two phases are also reflected in the number of components; high in the first phase, quickly becoming a single component in the second. In the null model, a similar effect occurs, though average shortest path length stays generally lower because of the lack of a connection preference such as the clustering mechanism in the range model.

As a direct result of the large difference in clustering and average shortest path length, the range model networks have much higher small-world index when range is low. Introducing spatial constraints produces more 'small-world-like' networks, most notably so when $r:g=0.2$.

\subsection{Range constrains the influence of population density on network structure.}

\begin{figure}
\includegraphics[width=\linewidth]{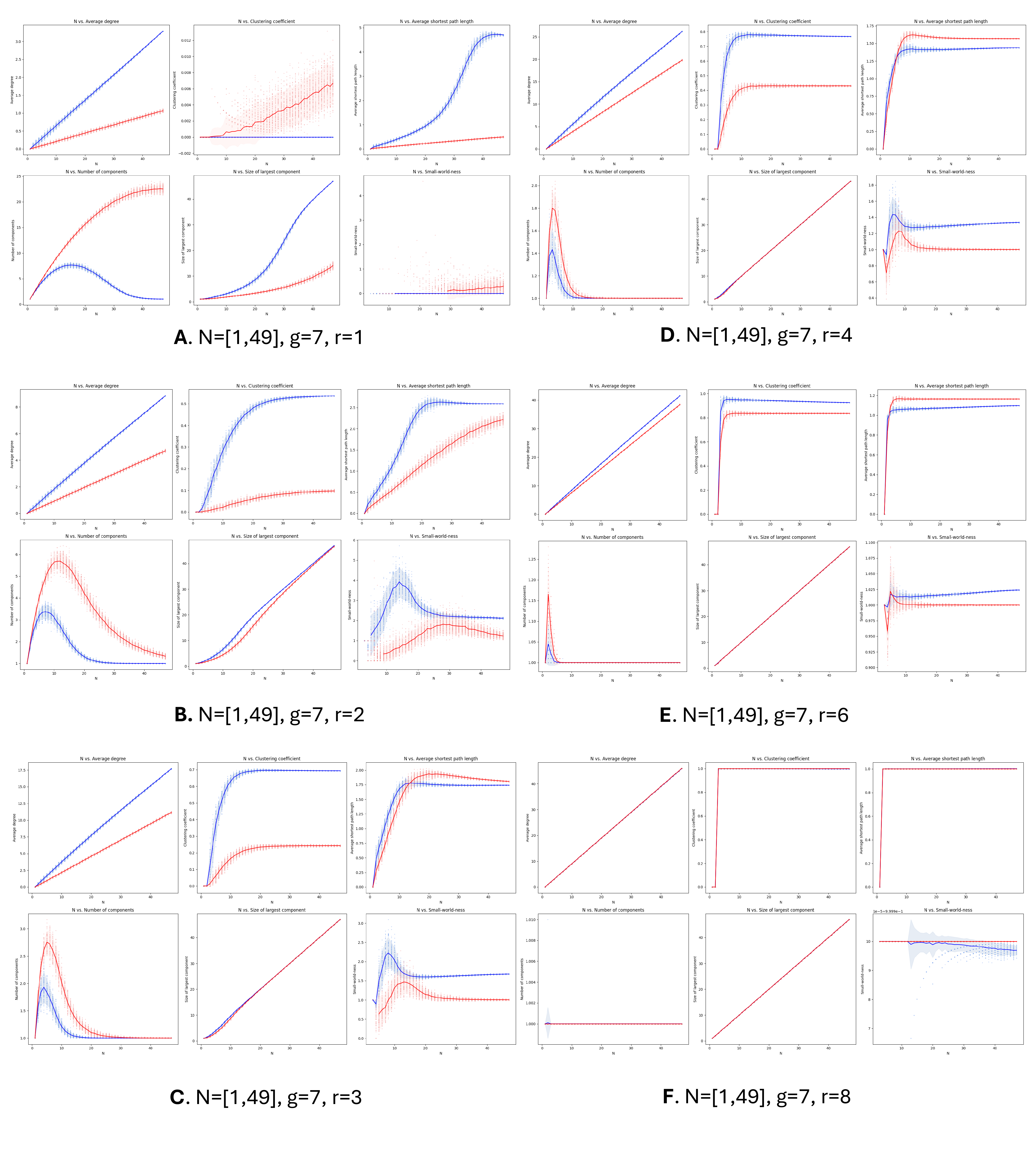}
\caption{{\bf Network properties of range model (blue) and null model (red) networks for varying \emph{N} between 1 and 49 (x-axes) within runs and varying \emph{r} 1, 2, 3, 4, 6, 8 between runs}. Dots in the graphs are time-averages of 100 timesteps, of which 100 rounds are collected at each input setting of \emph{r}. Lines are drawn through the means of these time averages, and transparent areas show 1.5 standard deviations. Varying \emph{r} has a clear impact on the relationship between population density and all 6 network measures, determining how densely connected graphs can become as \emph{N} increases. When \emph{r} approaches \emph{g}, the role of agent locations diminishes until networks in the range model become like the non-spatial null model.}
\label{fig4}
\end{figure}

For constant values of \emph{r} and \emph{g}, varying population size \emph{N} also impacts network structures in the range model. In both the null model and the range model, as population size increases, so do average degree and the size of the largest connected component. In the null model, degree increases with N because as more nodes are added, there are more possible links, meaning more instances of checking \emph{P(connect)} and therefore a higher probability of more connections per node. In both models, the size of the largest component simply increases with \emph{N} since the total number of nodes is higher. 

In the null model, average shortest path length and number of components show a slightly more complex relationship with \emph{N}.  For static Erdos-Renyi random graphs, it has been shown that for each connection probability value there is a critical \emph{N} at which a giant component is formed, including most or all nodes \citep{erdos1959random}. This dynamic version of the same random process shows a similar behavior. At lower values of \emph{N}, the average number of connected components in null model networks increases steeply. After reaching this peak, the number of connected components decreases again, reaching 2 at $N=g^2$. Average shortest path length follows a similar two-step progression, initially increasing, reaching a brief plateau, and then increasing again until reaching 2 at $N=g^2$. These two measures describe the formation of a giant component as \emph{N} increases. In the first phase of increasing \emph{N}, the number of components increases, as the connection chance plus the number of nodes 'trying' to connect is not high enough to form more than isolated smaller components of 1 or 2 nodes. Average shortest path length increases to a value between 0, where no nodes are connected, and 1, while the number of components stays at close to $N/2$, indicating that mostly connected pairs form. As \emph{N} increases further, nodes become more connected and the number of components starts to fall, tending towards a single component. Average shortest path length starts another slight increase because, while the network tends towards a single component, as more nodes are added, the network diameter (the longest shortest path in a network) of that component increases as well, increasing the probability that the distance between peripheral nodes and nodes at the opposite side of the component grows. Overall, however, average shortest path length stays low, reaching a maximum of 2, reflecting the ability of random graphs to mix well.

The range model echoes some of these dynamics, but adds others.  In the range model, the increase in degree and average shortest path length with \emph{N} is much steeper, and the number of connected component stays lower throughout. These additional effects of population size in the range model are due to agents being situated in a bounded space. As \emph{N} increases while \emph{g} remains constant (or when \emph{g} decreases while \emph{N} remains constant), population density $N/g x g$ increases. With population density, the probability of randomly moving agents being in range of each other increases, increasing the network density more than \emph{N} otherwise would, as expressed by a higher average degree and a lower number of connected components than in the null model. Because agents avoid shared coordinates, as \emph{N} increases they become both closer and more evenly spread out over the coordinate space, leading to single connected component for much lower values of \emph{N} than is the case in the null model, where random connections can leave nodes isolated. The additional increase of average shortest path length with \emph{N} is due to the effect of range-constrained communication described above, which leads to more clustered networks. 

For varying \emph{N} even more than for varying \emph{r}, the range model shows much higher clustering than the null model. Unlike average shortest path length, number of connected components, size of largest component, and degree, there is little to no relation between \\emph{N} and clustering in the null model. Once again this is consistent with existing knowledge of static random graphs: As \emph{N} increases, the number of possible edges grows, but the likelihood of any specific pair or triangle of nodes being connected remains governed by \emph{P(connect)} \citep{newman2003structure}. In the range model, by contrast, clustering increases with population density. As discussed above, the constraint on communication range tends to cluster nodes together. Additionally, graph density increases as population density increases, leading to an even bigger contrast with the null model.

Finally as \emph{N} increases, clustering, average shortest path length, and number of connected components reach a "plateau", remaining at a constant value. These are limits imposed by the range parameter. As population density increases, clustering would increase further if \emph{r} would not impose a limit on the possible connections that can be formed, and therefore on the number of connected triangles. Fig. \ref{fig4} visualizes how for different values of the range parameter, clustering reaches the plateau at different values. The same is true for average shortest path length: when range approaches \emph{g}, there are no restrictions on possible connections and the average shortest path length tends to 1, indicating a fully connected network. However, when range is lower, increasing \emph{N} increases the average shortest path length as population density increases the graph density while the network remains relatively clustered. As range decreases, the maximum average shortest path length reached increases. Note that this limit is also visible in the results for average degree. As explained above, a fully connected network has an average degree of $N-1$. In contrast to that, see for example Fig.4B: for $N=20$, maximum degree would be 19, but the results show an average degree of 3.9. Similarly low scores are shown throughout, degree staying far from the possible maximum. This, too, is due to range. Average degree does increase as more agents occupy the coordinate space, but because connections are limited to agents in range, the extent to which it increases is limited as well. As this limiting factor of range decreases as range approaches \emph{g}, the null and range models start to resemble each other more (see Fig. \ref{fig4}).

To summarize the simulation results, we found two major network formation mechanisms introduced by a coordinate space, random movement, and ranged communication. First, for any constant population size, the ratio of \emph{r} to \emph{g} determines how densely connected graphs are, from disconnected nodes at $r=0$ to fully connected networks when $r>g$. Second, independently from communication range, increasing the population density $N/g^2$ also increases graph density, as agents 'crowd' the coordinate space, making connections more likely, until they each occupy one of the possible coordinates when $N=g^2$. The impact of increasing population density on graph structures is still determined by \emph{r}, each node relying on movement for connection when population density is low, but reaching the maximum degree and clustering possible under the current value of \emph{r} when the population density reaches 1. Range (relative to coordinate space size \emph{g}) therefore sets the slope of the increase of graph density with population density, as well as 'plateau' maximum values to the increase in clustering, decrease in average shortest path length, and decrease in number of connected components.

In addition to tracking network properties outputs, our simulations included four commonly modelled processes of information diffusion on the networks. These simulations reproduce known results on network properties and information diffusion \citep{wattsCollectiveDynamicsSmallworld1998, barabasi2000scale, lazerNetworkStructureExploration2007, centolaNetworkScienceCollective2022}.  For example, when range is low and therefore the number of connected components and clustering are high, information spreads more slowly, and the population explores more complex problem spaces more effectively than the null model. These results illustrate how spatial constraints impact not just our predictions of communication networks, but also (and as a result) our predictions of transmission processes that take place on these networks. For details, see the supplemental materials.

\section{Discussion}
In this paper, we report simulations demonstrating that including spatial constraints on communication introduces significant differences in network architectures (and by extension in information flow: see supplemental materials). In general but especially when range is low compared to environment size \emph{g}, networks are more clustered, have higher average shortest path length, a lower number of connected components, and a higher small-world index than dynamic random graphs. Two main parameters drive average network structures: communication range \emph{r}, relative to \emph{g}, and population density $N/g^2$. As either increases while the other remains constant, graphs become more densely connected. The influence of population density on graph density is bounded by \emph{r}, limiting the possible connections agents can make. As population density increases, movement becomes increasingly impossible as every agent occupies one of the $g*g$ coordinates when $N=g^2$. As a result, positions become fixed and node properties become global, set by the value of \emph{r}. A specific mechanism that causes higher clustering and average shortest path lengths in ranged networks is that any agents \emph{j} connecting to an agent \emph{i} need to be within the area covered by \emph{i}'s range, significantly increasing the probability that agents \emph{j} connect to each other as well. We also show that range-constrained networks create more small-world like networks. 

One obvious but central result of our simulations is that when communication range \emph{r} exceeds grid size \emph{g}, fully connected communication networks are guaranteed. In other words, $r>g$ is the condition that needs to obtain before each individual has communicative access to each other individual. By extension, this is the condition that needs to obtain before individual choices, strategies, and biases about who to connect to and communicate with can fully account for network formation. Our simulations quantify what happens when this condition does not obtain (when $0<r<g$), and when therefore spatial constraints play a role in network formation. An interesting direction for future work, in models and experiments, would be to combine individual connection biases with spatial constraints. 

One mechanism that starts to play a role in network formation when $r<g$ is mobility. In their paper reporting a similar model, Vining and colleagues discuss a distinction between solid and liquid brains \citep{viningHowDoesMobility2019}.  Solid brains, such as human brains and computer chips, have elements and links in more or less fixed positions, whereas liquid brains, such as swarms and slime moulds, consist of mobile elements that form transient links \citep{soleLiquidBrainsSolid2019}. Vining and collegues note that in liquid brains, mobility can play the role that physical links can play otherwise: connecting nodes, creating a pathway enabling information sharing. This is what we see in our simulations when $r>g$: because agents are situated in an environment and communication is range-constrained, moving through that environment becomes a mechanism for network formation. As $r:g$ decreases, mobility becomes the dominant mechanism for connection; as $r:g$ increases, links take over. Additionally, as population density increases, node properties becomes increasingly global and determined by \emph{r} as the coordinate space becomes saturated and mobility becomes impossible. As the mobility mechanism disappears, models of networks on fixed lattices may therefore become more applicable under these circumstances. 

Real individuals will of course show more intelligent mobility patterns than random movement. Moreover, random movement can be implemented in ways different from our model. A more standard implementation of movement direction in agent-based models is for agents to select from a uniform distribution of angles rather than directly from integer coordinate 'tiles.' In a recent paper, Chimento and Farine simulated this implementation of random movement as well as movement patterns where the distribution of angles was more concentrated towards specific directions, and where agents move towards resources in the environment \citep{chimentoContributionMovementSocial2024}. Using data from cell phone towers, researchers have found that human displacements over time are approximated by agents selecting from a power-law distribution of distances. Most of us move small to no distances most of the time, and greater distances some of the time; in addition, a small number of people travels much greater distances \citep{gonzalezUnderstandingIndividualHuman2008, almaatouqMobileCommunicationSignatures2016}. One direction for future work is to combine our results of varying range and population size with these more realistic mobility patterns, and to test the fit of different mobility patterns to observational data of human communication networks, employing our model to see the role of spatial constraints to communication in the real world.

Apart from mobility, as discussed, we found a role for population size and density in the formation of communication networks. The role of population size in transmission processes has been discussed in literature on models in cultural evolution. Models have demonstrated that greater population size, under the right demographic and network connectivity circumstances, increases a population's ability to create and maintain complex skills \citep{moserInnovationfacilitatingNetworksCreate, baldiniRevisitingEffectPopulation, deffnerEffectivePopulationSize2021, henrichDemographyCulturalEvolution2004}. Our simulations show a different role for population size: inreasing $N$ in a bounded space increases population density, increasing the probability for randomly moving agents to encounter each other, creating denser networks for communication. Under the spatial constraints of a bounded environment and limited-range communication, population size influences not just transmission, but the communication network infrastructure that makes transmission possible.

One limitation of the present work is that though we quantify the impact of a bounded environment, the space itself is empty and unobstructed. The structure of real environments might significantly impact the possiblities of communication, introducing more complexity as well as realism into the spatial computation of network structure. A second limitation is that we consider only single-layer networks with global values for communication range, while real communicating systems may use different and overlapping communication media and technologies with different effective ranges. A third limitation is that we compare the ranged model to only one type of non-spatial model. Comparison to (dynamic versions of) other 'standard' non-spatial network generation algorithms such as small-world and connected-caveman graphs \citep{wattsCollectiveDynamicsSmallworld1998, watts1999networks} would paint a broader picture of the impact of including spatial constraints in network generation algorithms. All three of these limitations would be fruitful directions for future work.

Leaving the role of spatial constraints out of the picture (for example, by over-relying on digital communication data, experimental settings where participants are all in the same space, or traditional network generation algorithms) could lead one to suggest that the mechanisms driving human communication network formation are primarily in psychology and behavior. However, including them in a model shows that, compared to a non-spatial model, these constraints alone can have a major impact on our predictions of network structure. One advantage of physical properties such as location and communication range as an explanatory mechanism of real communication networks is that they are much more straightforward to interpret, test and implement than hypotheses about drivers of human psychology and behavior. The simulations reported in this paper provide tools to understand communication networks as the product of a process of collective computation which does not just include choices by isolated individuals but also their surrounding environment.  

\bibliographystyle{plos2015}
\bibliography{References}

\end{document}


\title{Supplemental materials for "Connect-while-in-range: modelling the impact of spatial constraints on dynamic communication network structures"}

\maketitle

\section{Model code}
This model was coded in Python using the Networkx package. The code including detailed instructions is publically available on github: https://github.com/niekkerssies/Range-model.

\section{Varying g}
In the article, \emph{g} is the only one of the three model parameters that isn't varied across its complete meaningful range (we report results on $g=10$ for varying \emph{r}, and $g=7$ for varying \emph{N}). This is because there is no analogue of \emph{g} in the null model. The parameter is hard to compare to non-spatial models, and in either case shows up mainly by its influence on population density. The behavior of the \emph{g} parameter, however, does merit its own report. Fig. S1 shows results for varying \emph{g} between 3 and 50 for $N=9$.

\begin{figure}[!h]
\includegraphics[width=\linewidth]{Results g final.png}
\caption{{\bf }}
\label{figS1}
\end{figure}

The results of increasing \emph{g} for constant \emph{N} on network structure are generally the inverse of the effects of increasing \emph{N} for constant \emph{g}. For very low \emph{g}, where there is a low number of total possible coordinates, population density will be equal to 1 or close to 1 (depending on \emph{N}), creating fully connected graphs as long as $r\geq g$. As \emph{g} increases, graph density decreases (average degree and largest component decrease; number of components increases). The major difference from the effect of \emph{N} on network structure is the way \emph{g} scales: since the grid is defined as \emph{g*g}, the coordinate space scales quadratically, while \emph{N} scales linearly. This is visible, for example, in the curve of degree with increasing \emph{g} (Fig. S1) compared to the straight line of degree with increasing \emph{N} (Fig. 4 main article).

\section{Transmission processes}
On top of the communication networks that result from the range and null model, four models of diffusion processes were simulated: SI diffusion, complex contagion, cultural evolution, and a potion task. 

In SI transmission, a standard approach in epidemiology but often used for simulating information diffusion, agents can be either Susceptible (S) or Infected (I). In the simulation, a number \emph{N\_init} of agents are initialized in state I, and the rest in state S. Each timestep, after movement and link updating (or in the non-spatial model, just link updating) has occured, each agent checks the state of their network neighbors. If an S-agent is connected to an I-agent, the S-agent becomes infected with probability \emph{P(infect)}. 

Complex contagion is defined in contrast to "simple contagion,"  of which SI transmission is an example \citep{centolaComplexContagionsWeakness2007}. In simple contagion, S-state individuals are infected by exposure to I-state individuals with some fixed global probability \emph{P(infect)}. In complex contagion, infection probability depends on the number of infected neighbors. Specifically, in the simulation, an S-state agent \emph{i} connected to neighbors \emph{j1,j2,....,jn}  with probability $P +( Ij / N)w$, where \emph{P} is the baseline infection probability, \emph{Ij} is the number of infected neighbors, \emph{N} is the population size, and \emph{w} is a social weight value.  As the number of infected neighbors increases (relative to the total population), it becomes more likely for \emph{i} to be infected. 

Another model of a collective information process is cultural evolution. Models of cultural evolution are based on population genetics models that track the intergenerational transmission of genetic variants, for example two variants 'A' and 'B'. In contrast to the SI model, where agents can be in a infectious state with a transmission probability, and a susceptible state, agents can now be in two states that each have their own transmission probability, pA and pB. When these probabilities are the same, this is called unbiased transmission or "cultural drift." As is the case in genetic drift, the information that survives until the next generation can be an imperfect sampling of the previous generation, overrepresenting one trait over another not by selective pressure but by chance. When the transmission probabilities are unequal and one variant becomes therefore more likely to be transmitted, the model is called biased transmission \citep{acerbiMesoudiSmollaIndividualBasedModels2022}.

The potion task, based on an experiment \citep{derexPartialConnectivityIncreases2016} and recreated in several existing models \citep{miglianoHuntergathererMultilevelSociality2020a, moserInnovationfacilitatingNetworksCreate, derexDivideConquerIntermediate2018},  is more specific. Participants were placed in groups, started with an inventory of items ('ingredients') associated with scores, and could see items and known combinations ('potions') from other group members. Some items combined to create items with higher scores (for example, items a1, a3 and b2 combined to make B1). These higher-level items make other valid combinations, creating a 'trajectory' towards increasingly complex items. The valid combinations were chosen so that two such trajectories existed, but the most complex combination in both trajectories included both A3 and B3, the next-most complex items from both trajectories. The original experiment found that more sparsely connected groups are succesful at exploring both trajectories and therefore finding the final item, while fully connected groups usually only optimized within one trajectory.

In this model, the potion task is simulated as follows. Agents are initialized with a starting inventory, as specified by the original experiment. Then, they select a random network neighbor, and the two agents randomly select 3 items from their inventories (at random, one agent picks one or two, and the second agent picks the remaining number). Higher-scored items are more likely to be selected. Then, if this is a valid combination, both agents add the combination product item to their inventory. If a new valid combination was found, each of the two agents involved diffuse this new item to their network neighbors with probability \emph{p(diff)}.

\subsection{Test setup}
For each varying parameter, one test was done for each of the 4 transmission processes. To track SI diffusion and complex contagion, the frequency of agents in the I-state (a value between 0 and 1, where 0 indicates no infected agents and 1 indicates the entire population is infected) was collected each timestep. If frequency reached 1, the timestep was collected at which it did (fixation time). This gives an indication of the speed of diffusion along the network. The same measures were collected for both traits 'A' and 'B' in the cultural evolution model. In the potion task model, rather than fixation times, the interest is in "crossover events." \citep{derexPartialConnectivityIncreases2016}. For both trajectories, there are 3 tiers of valid combinations that lead to increasingly higher scores, and a fourth combination that requires the tier-3 potions from both trajectories. If a 4th-tier item is created, the timestep was collected at which it did. Whether a crossover was reached is an indication of the population exploring the full problem space, and the time at which it was reached gives an indication of how efficiently it is able to do so. 

\begin{figure}[!h]
	\includegraphics[width=\linewidth]{Diffusion r.jpg}
	\caption{Trait frequencies over time (SI diffusion, top; complex contagion, second from top; biased transmission, third from top) for $N=10$, $g=10$; and potion task crossover events out of 100 timesteps for $N=80$, as range varies between 0 and 10. In the biased transmission model, negative values of the trait frequency indicate a dominance of trait B, while positive values indicate A. Overall and especially at lower range values, SI diffusion is slower and complex contagion and biased transmission are faster in the range model networks. In both models, diffusion speeds up as networks become more connected. Potion task performance is highest in intermediate values of range and connection probability, and the null model scores higher in overall performance.}
\label{figS2}
\end{figure}

\begin{figure}[!h]
	\includegraphics[width=\linewidth]{Diffusion N.pdf}
	\caption{Trait frequencies over time (SI diffusion, top; complex contagion, second from top; biased transmission, third from top) for $r=2$, $g=10$; and potion task crossover events out of 100 timesteps, \emph{N} varying between 1 and 24. Transmission processes show the same between-model differences as under varying \emph{g}. Transmission overall slows down as \emph{N} increases and the traits have to reach a larger population. Potion task performance clearly scales with \emph{N} in both models.}
\label{figS3}
\end{figure}

\subsection{Results}
According to existing literature, the differences in communication network structures introduced by spatial constraints should lead to different predictions about information diffusion taking place on the networks. Since the range model networks have, in general, a higher clustering coefficient and higher average shortest path length, information modelled as the I-state in a SI diffusion process should spread more slowly than in the random networks of the null model \citep{watts1999networks}. For a complex contagion process, the opposite is true: because clusters facilitate more redundant reinforcement by local connections, increasing the transmission probability, much higher clustering should mean faster diffusion through the range model networks \citep{centolaComplexContagionsWeakness2007}. In a potion task, firstly, the original experiment as well as follow-up models found that sparsely connected networks perform better at the task than dense or fully connected networks (they find more 'crossover' items)\citep{derexPartialConnectivityIncreases2016, moserInnovationfacilitatingNetworksCreate, derexDivideConquerIntermediate2018}. Secondly, increasing population size was found to increase performance a result that can be accounted for simply by the fact that more combinations are tried (though see \citep{baldiniRevisitingEffectPopulation}). 

All of these known effects are reproduced here, leading to differences between the spatial and non-spatial model in all four diffusion processes modelled. SI diffusion is slower in range model networks for all values of \emph{r} and \emph{P(connect)}. The differences are most notable when range is low, which is also when network structure differences are most pronounced. Complex contagion shows the opposite: overall and especially at lower range values, diffusion is faster in range model networks. For both of these processes, diffusion eventually becomes much slower when population density decreases to the point where range networks become disconnected. Interestingly, for a 'biased transmission' cultural evolution process where transmission probability of trait A $pA = 0.1$, and $pB = 0.2$, the population reaches fixation (all agents adopt the same trait) of trait B slightly faster than in the null model, once again especially at lower values for range. 

In the potion task, the effects of both population size and graph density were replicated. As population size increases, populations in both the range and the null model produce more crossover events, of which most occured between $r:g=0.1$ and $r:g=0.4$, and $P(connect)=0.2$ and $P(connect=0.5)$. As expected, the range model performs better than the null model at low range, when graphs are much more clustered and diffusion slower (see Fig. S2D). However, overall, the random model performed better, finding close to 10$\%$ more crossovers across settings. This is consistent with the fact that the null model creates a higher number of connected components. Earlier results point to the fact that in this specific task, it's not just the intermediate to low graph density that increases performance, but specifically the number of cliques \citep{moserInnovationfacilitatingNetworksCreate}. Cliques are a term for more connected node communities within connected graphs. Strictly speaking, looking at the static graphs of single timesteps, the range model exhibits more mutually disconnected components than cliques. However, in this dynamic simulation, connected components can function as cliques over time, as they can become connected when links change. In a cumulative represenation of this dynamic process, groups of nodes that form isolated components most of the time and connected cliques some of the time will show up as connected cliques. 

These diffusion processes are hypothetical and involve assumptions about how individuals receive and transmit information that are hard to verify or falsify. Nevertheless, the results demonstrate that including spatial constraints on the communication networks on which information diffusion happens leads to differences in prediction in four common models of diffusion. This amounts to an argument to include spatial constraints not just in models of network formation, but in models of diffusion as well; it matters for prediction of diffusion outcomes.

\newpage
\section{Varying over larger N}

\begin{figure}[!h]
	\includegraphics[width=\linewidth]{Varying N 100 range 2.png}
	\caption{{\bf Network properties of range model (blue) and null model (red) networks for varying \emph{N} between 1 and 100 (x-axes), $g=10$, at $r=2$.} The behavior of network structure properties is robust to higher \emph{N} than shown in the article. Compare this figure to Fig. 4B, where \emph{N} varies over a lower range but \emph{r} is set to the same value (2). The absolute values are different, but the response of network structures to increasing \emph{N} shows the same pattern.}
\label{figS4}
\end{figure}

\bibliographystyle{plos2015}
\bibliography{References}